\documentclass[fleqn,12pt,twoside]{article}
\usepackage[headings]{espcrc1}

\readRCS
$Id: espcrc1.tex,v 1.2 2004/02/24 11:22:11 spepping Exp $
\usepackage{graphicx}
\usepackage[figuresright]{rotating}

\newcommand{\AmS}{{\protect\the\textfont2
  A\kern-.1667em\lower.5ex\hbox{M}\kern-.125emS}}
\newcommand{\be}{\begin{eqnarray}}
\newcommand{\ee}{\end{eqnarray}}

\newcommand{\np}[1]{Nucl. Phys. { #1}}

\usepackage{graphicx}

\title {\bf Unbound exotic nuclei studied by projectile fragmentation \footnote{\uppercase {I}n memory of \uppercase{A}delchi \uppercase{F}abrocini.}}

\author{Guillaume Blanchon\address {\it Scuola di Dottorato G. Galilei, Pisa.}\address[infn] {\it INFN, Sez. di Pisa and
 Dipartimento di Fisica, Universit\`a di Pisa,   Largo Pontecorvo 3, 56127 Pisa, Italy.},  A. Bonaccorso\addressmark,  D. M. Brink\address {\it Department of Theoretical Physics, 1 Keble Road, Oxford  
OX1 3NP, U. K.}, A. Garc\'ia- Camacho\addressmark[infn] and$~$ \\N. Vinh Mau\address {\it Institut de Physique Nucl\'eaire, IN2P3-CNRS, F-91406,  Orsay Cedex, France.}.}
\begin{document}

\maketitle

\begin{abstract}
We call {\it projectile fragmentation} of neutron halo nuclei  the  elastic breakup (diffraction) reaction, when the observable studied is the neutron-core relative energy spectrum. This observable has been  measured in relation to the Coulomb breakup on heavy target and recently also on light targets. Such data enlighten the effect of the neutron final state interaction with the core of origin. Projectile fragmentation is studied here by a  time dependent model for the excitation of a  
nucleon from a bound state to a continuum resonant state in a
neutron-core complex potential which acts as a final state interaction.  
   The
final state is described by an optical model S-matrix so that both  
resonant and non resonant states of any continuum energy can be studied as well
as deeply bound initial states. It turns out that due to the coupling  
between the initial and final states, the neutron-core free
particle phase shifts are modified, in the exit channel, by an  
additional phase. 
 Some typical numerical calculations for  the relevant observables are  presented and compared to experimental data.  It is suggest that the excitation energy spectra of an  
unbound nucleus might reflect the structure of the parent nucleus from
whose fragmentation they are obtained.\end{abstract}

\section{Introduction}

All theoretical methods used so far to describe breakup rely on a basic approximation to describe the collision with
only the three-body variables of nucleon coordinate, projectile
coordinate, and target coordinate. Thus the dynamics is controlled by the
three potentials describing nucleon-core, nucleon-target, and core-target
interactions.  In most cases the projectile-target relative motion is treated
semiclassically by using a trajectory of the center of the projectile
relative to the center of the target $\mathbf{R}%
\left(  t\right)  =\mathbf{b}_c+vt\mathbf{\hat{z}}$ with constant velocity $v$ in the $z$ direction and impact parameter
{\bf b$_c$} in the $xy$ plane.  This approximation makes our formalism applicable for incident energies above the Coulomb barrier. Along this trajectory the amplitude for a
transition from a nucleon state  $\psi_i$  bound  in the projectile, to a final  continuum state $\psi_f$, is given by \cite{adl,bb}
\begin{equation}A_{fi}={1\over i\hbar}\int_{-\infty}^{\infty}dt \langle \psi_{f} ({\bf r},t)|V({\bf { r, R}}(t))|\psi_{i}({\bf r},t)\rangle,\label{1}
\end{equation}
where $V$ is the
interaction responsible for the transition which will be specified in the following. The probabilities for different processes can be
represented in terms of the amplitude as ${dP/d
\xi}=\sum |A_{fi}|^2 \delta(\xi-\xi_f)\label{2}$ where
$\xi$ can be momentum, energy or any other variable for which a differential cross section is measured.
Direct one-particle re-arrangment reactions of the peripheral type in presence of strong core-target absorption can be described by an equation like  \cite{bb,bb4,1,bw}
\begin{equation}
{d\sigma_{-n}\over {d{{\xi_f}} }}=C^2S
\int d{\bf b_c} {d P_{b_{up}}(b_c)\over
d{{\xi_f}}}
P_{ct}(b_c), \label{cross}
\end{equation}
(see Eq. (2.3) of \cite{bb4}) and C$^2$S is the spectroscopic factor  
for the initial single particle state.
The core survival probability is defined in terms
of a S-matrix function of the core-target distance of closest approach  
$b_c$. A simple parameterisation is
 $P_{ct}(b_c)=|S_{ct}|^2=
e^{(-\ln 2 exp[(R_s-b_c)/a])}.$  
It  takes into account the peripheral nature of
the reaction and naturally excludes the possibility of large overlaps  
between projectile and target. The  
strong absorption radius R$_s\approx1.4(A_p^{1/3}+A_t^{1/3})$ fm is  
defined
as the distance of closest approach for a trajectory that is 50\%  
absorbed from the elastic channel and a=0.6 fm  is a diffuseness
parameter. The values of R$_s$ thus obtained agree within a few percent  
with those of the Kox parameterization\cite{kox}.

\section{Projectile Fragmentation}
Let us call {\it projectile fragmentation}  the  elastic breakup (diffraction dissociation) reaction, when the observable studied is the neutron-core relative energy spectrum. This kind of observable has been widely  measured in relation to the Coulomb breakup on heavy target. Results on light targets have also been presented  \cite{fuku}. These data enlighten the effect of the neutron final state interaction with the core of origin, while observables like the core energy or momentum distributions enlighten the effect of the neutron final state interaction with the target.

Projectile fragmentation has also been used to study  two neutron halo projectiles \cite{nig}-\cite{Labi99}. In this case  it has been suggested that the  reaction might proceed by  
the simultaneous emission of the two neutrons or by
successive emissions \cite{nig}. The successive emission can be due to a mechanism in which one neutron is stripped by  the interaction with the target, as in the one-neutron fragmentation case, while the other is left behind, for example in a resonance state, which then decays. This mechanism has been described by the sudden approximation\cite{bhe} under the hypothesis that while the first neutron is stripped, the second neutron is emitted at large impact parameters with no final state interaction with the target. The emission can be expected sequential if the two neutrons are not strongly correlated.

If the two neutrons are strongly correlated they will preferentially be emitted simultaneously.  If  the neutron which is not detected is stripped while the other suffers an elastic scattering on the target, then
in both cases to first order in the interaction the neutron  ends-up in a plane wave final state \cite{bb}. It can then re-interact with the core which, for example,  is going to be $^{10}$Be in the case of the one-neutron halo projectile $^{11}$Be, while it will be  $^{12}$Be in the case of the projectile fragmentation of $^{14}$Be, since $^{13}$Be is not bound. 
Experiments with a  $^{14}$B projectile \cite{jl} have also been performed, in which the n-$^{12}$Be relative energy spectra have been reconstructed by coincidence  measurements. In such a nucleus the valence neutron is weakly bound, while the valence proton is strongly bound. Thus the neutron will probably be emitted in the first step and then re-scattered by the core minus one proton nucleus. The projectile-target distances at which this kind of mechanism would be relevant are probably not so large to neglect the effect of the neutron-target interaction.  
 \subsection {Inelastic excitation to the continuum. }

To first order the inelastic-like excitations can be described again by the  
time dependent
perturbation amplitude Eq.(\ref{1}) \cite{adl,bb}.
 In this section also, the potential $V({\bf { r,R}}(t))$, which  is the interaction responsible for the neutron transition,
 moves past on a constant velocity path as described in the previous sections.     The  radial part $ \phi_{i}({\bf r})$  of  the single particle initial state wave function $\psi_{i} ({\bf r},t)$ is calculated in a potential $V_{WS}( r)$  which is  
fixed in space.
The coordinate system  and other details  of the calculations can be found in  Ref.\cite{5}.  
 In the special case of exotic nuclei
the traditional approach to inelastic excitations needs to be modified.    For example the final state can be eigenstate of a potential $V_1$
modified with respect to $V_{WS}$  because some other particle is  
emitted during the reaction process as discussed in the introduction. The final state interaction might also have  
an imaginary part which would take into account the  
coupling between
a continuum state and an excited core. 
The first order time dependent perturbation amplitude then reads
\begin{equation}
A_{fi}={1\over i\hbar v}
\int_{-\infty}^{\infty}dx dy dz ~ \phi^*_{f}
(x,y,z)\phi_{i}(x,y,z){e^{iqz}}\tilde V(x-b_c,y,q),\label{c}
\end{equation}
where
$\tilde V(x-b_c,y,q)= \int_{-\infty}^{\infty}dz V(x-b_c,y,z)e^{iqz},$ and we changed  variables and put $z^{\prime}=z - vt$ or $t = (z -  
z^{\prime})/v$,
$
q={{\varepsilon_f-\varepsilon_i}/ {\hbar v}}
$. Here $\varepsilon_f$ is the neutron-core relative energy in the final state.

The r\^ole of the target represented by $\tilde V$ is just to  perturb the initial bound state wave function and to allow the transition to the continuum by transferring some momentum to the neutron. Then it is enough to choose a simplified form of the interaction,  such as a delta-function potential
$V(r) = v_2\delta(x)\delta(y)\delta(z)$. The value of the strength  $v_2\equiv$ [MeV fm$^3$]
 used in the calculation is taken equal to the volume integral of the appropriate neutron-target interaction. It is clear that  while in the sudden approach the initial and final state overlap is taken in the whole coordinate space, irrespective of the target and of the beam velocity, here 
the overlap of the initial and final wave functions depends on the core-target impact parameter. The neutron is emitted preferentially on the reaction plane  and the z-component, being along the relative velocity axis is boosted by a momentum $q$.

Due of the strong core absorption discussed in Sec.1
these calculations are  performed using the asymptotic form of the initial and final state wave functions. 
Introducing the quantization condition\cite{bb}
the probability spectrum reads
\begin{eqnarray}
{dP_{in}\over d\varepsilon_f}={2\over \pi}{v_2^2\over \hbar^2  
v^2}{C_i^2 }{m\over\hbar^2k}{1\over 2l_i+1}\Sigma_{m_i,m_f}
 |1-\bar S_{m_i,m_f}|^2 |I_{m_i,m_f}|^2. \label{8}
\end{eqnarray}
 The generalization including spin is given in Appendix B of Ref.\cite{5} and   $  |I_{m_i,m_f}|^2=\left|\int_{-\infty}^{\infty} dze^{iqz}i^{l_i} \gamma h^{(1)}_{l_i}(i\gamma  
r)Y_{{l_i},{m_i}}(\theta,0)  k{i\over 2}h^{(-)}_{l_f}(kr)Y_{{l_f},{m_f}}(\theta,0)\right|^2.$
The quantity $\bar S=e^{2i(\delta+\nu)}$ is an off-the-energy-shell S-matrix representing the final state interaction  
of the neutron with the projectile core.  It depends on a phase which is the sum of $\delta$, the free particle n-core phase shift, plus  $\nu$ the phase of the matrix element  $|I|$.

\section{Applications }
\subsection{The reaction $^{11}$Be $\to$ n+$^{10}$Be}

\vskip 12pt
\begin{figure}[htb]\begin {center}
 \includegraphics [scale=.25,angle=0]{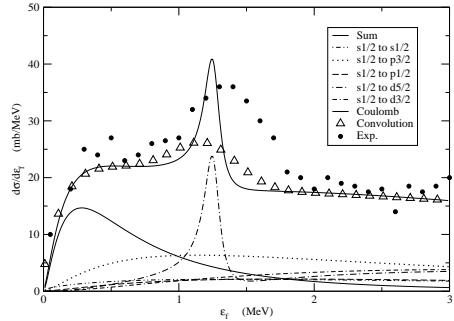}
\caption { \footnotesize n-$^{10}$Be relative energy  spectrum,
including Coulomb and nuclear breakup for the reaction  $^{11}$Be+$^{12}$C $\to$ n+$^{10}$Be+X at 69 A.MeV. Only the contributions from an s initial state with  
spectroscopic factor C$^2$S= 0.84 are  calculated. The  triangles are the total calculated result after convolution with the experimental resolution function. The dots are the experimental points from \cite{fuku}.  }
\end{center}
\label{fig1}
 \end{figure}

As a test of our model we calculate the relative energy spectrum  n+$^{10}$Be obtained by the authors of Ref.\cite{fuku}
in the breakup reaction of $^{11}$Be on $^{12}$C at 70 A.MeV. The structure of $^{11}$Be is  well known: the valence neutron is bound by 0.503 MeV; the wave function is mainly a 2s state with a spectroscopic factor around 0.8 and there is also a small d$_{5/2}$ component. The main d$_{5/2}$ strength is in the continuum centered around 1.25.  We have  calculated the initial wave function for the s-state in a simple Woods-Saxon potential with strength fitted to the experimental separation energy and whose parameters are: r$_0$=1.25 fm, a=0.8 fm. As possible final states we have considered only the s, p and d partial waves calculated in the $l$-dependent potentials of \cite{5}. The delta-function potential strength
has been chosen as -4057.59 MeV fm$^3$. The authors of Ref.\cite{fuku} have shown that the effect of Coulomb breakup is noticeable in  their  n+$^{10}$Be spectrum. We have also included this contribution, calculating it according to \cite{1}. 
The spectrum of Fig.\ref{fig1} is very similar to the spectrum obtained in Ref.\cite{cap} by solving 
the time-dependent Schr\"odinger equation  numerically,  expanding 
the projectile wave function upon a three-dimensional spherical mesh. Similarly to the present model, a classical, straight line trajectory for the core-target scattering was used in Ref.\cite{cap}. Also our n-core potentials are very close to those
of Ref.\cite{cap} and our $\delta$-interaction strength is consistent with the volume integral of their neutron-target interaction.
 We have then  folded the calculated spectrum througth the experimental resolution function of  Fukuda et al. \cite{fuku}, as given in Ref.\cite{cap}. The result is shown in Fig.\ref{fig1} by the triangles. The full curve is the total spectrum, sum of Coulomb and nuclear breakup. Each individual transition, due to the nuclear interaction only, is also shown.
The dots are the experimental points from \cite{fuku}. The kind of discrepancy between  our calculation and the data in the range 1-2 MeV is very similar to that of the calculations in Ref.\cite{cap}.

\subsection{ Structure of  $^{14}$Be and $^{14}$B }

Uncertainties in the interpretation of experimental results as  
compared to structure calculations were at the origin of our  
motivations to
try to understand  whether the neutron-$^{12}$Be relative energy  
spectra obtained
from fragmentation of $^{14}$Be or $^{14}$B  would show differences  
predictable in a theoretical model.
If differences will be found  in the experimental results
with $^{14}$B and $^{14}$Be beams they could be  due to an interplay between structure and reaction effects. 

The ground state of $^{14}$Be  
has spin $J^\pi=0^+$. In a simple model assuming two
neutrons added to a $^{12}$Be core in its ground state the wave  
function is:
\begin{equation}
|^{14}Be>=[b_1(2s_{1/2})^2+b_2(1p_{1/2})^2+b_3(1d_{5/ 
2})^2]\otimes|^{12}Be,0^+>.\label{l}
\end{equation}
Then the bound neutron can be in a 2s, 1p$_{1/2}$ or 1d$_{5/2}$ state.  
However, as it has been discussed in the previous section, the situation is much more complicated \cite{Bert91}-\cite{de} and  in particular the
calculations of Ref. \cite{ta} show that there is a large component $(2s_{1/2},1d_{5/ 
2})\otimes|^{12}Be,2^+>$
 with the core in  
its low energy 2$^+$
state which can modify the neutron distribution. 
\begin{figure}[ht!]\begin{center}
               \includegraphics[scale=.25,angle=0]{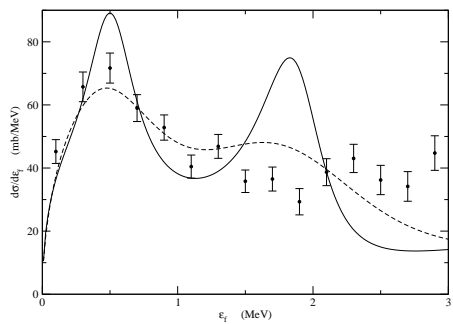}
\caption{ \footnotesize Sum  
of all transitions from the s initial state   for the reaction $^{14}$Be+$^{12}$C $\to$ n+$^{12}$Be+X . Experimental points from  \cite{simo}.  Dashed line is the folding of the calculated spectrum with the experimental resolution curve.}\end{center}
\label{fig2}
\end{figure}

The ground state of $^{14}$B has spin $J^\pi=2^-$. 
In a model where it is described as a neutron-proton pair
added to a $^{12}$Be core in its 0$^+$ground state with the proton in  
the 1p$_{3/2}$
shell, its wave function may be written as:
\begin{equation}
|^{14}B>=[a_1(p_{3/2},2s_{1/2})+a_2(p_{3/2},  
d_{5/2})]\otimes|^{12}Be,0^+>.\label{m}
\end{equation}
The present experimental information \cite{msu} on  $^{14}$B is that the
neutron is in a state combination of s and d-components with weights  
66\% and 30\% respectively, while shell model calculations show a similar
mixture and no component with an excited state of the core.
There are two possibilities for the reaction mechanism. One is that
a proton is knocked out in the reaction with the target. The remaining  
$^{13}$Be would be left
in an unbound  s-state with probability $| a_1 |^2$, in a  
d$_{5/2}$-state with probability
$| a_2 |^2 $.  These unbound
states would decay showing the s-wave threshold and d-wave resonance  
effects. As mentioned in the introduction, the second possibility is that the neutron is knocked out first due to its small separation energy and that the proton is stripped from the remaining  $^{13}$B. 

To give another example of a possible comparison with available  data, we show in Fig. \ref{fig2} the experimental points from H. Simon et al. \cite{simo} for the reaction $^{14}$Be+$^{12}$C $\to$ n+$^{12}$Be+X at 250 A.MeV. The normalization factor of the data  to mb/MeV is 0.843. The solid line gives the sum  
of all transitions from the s initial state with $\varepsilon_f$=-1.85 MeV (solid line), renormalized with a factor 2.4.  The dashed line is the folding of the calculated spectrum with the experimental resolution curve. Therefore the calculation underestimate the absolute experimental cross section by a factor of 2.   In view of the incertitude in the strength of our n-target $\delta$-potential and on the initial state spectroscopic factor which has been taken as unit, we can consider   our absolute cross sections  quite  reasonable. A more detailed account of these calculations is given in \cite{5}.

\section{Conclusions and Outlook}
The field of Rare Isotopes Studies is very active, growing steadily  and rapidly. Some recent achievements in the reaction theory for elastic breakup  have been presented.   From the structure point of view, in the search for the dripline position, a very important role is played by the study of nuclei unstable by neutron emission.  This is one of the most important subjects which need to be adressed and further developed in the near future and for which some suggestions have been presented.

\end{document}